\documentclass[runningheads]{llncs}
\usepackage{graphicx}
\usepackage{cite}  %
\usepackage{letltxmacro}  %
\usepackage{amssymb}%
\usepackage{amsmath}%
\usepackage{verbatim}
\usepackage{color}%
\usepackage{eso-pic}
\newcommand\AtPageUpperMyright[1]{\AtPageUpperLeft{%
 \put(\LenToUnit{0.3\paperwidth},\LenToUnit{-1.5cm}){%
     \parbox{0.8\textwidth}{\raggedleft\fontsize{9}{11}\selectfont #1}}%
 }}%
\newcommand{\conf}[1]{%
\AddToShipoutPictureBG*{%
\AtPageUpperMyright{#1}
}
}

\begin{document}
\title{Brain tumor segmentation with missing modalities via latent multi-source correlation representation}
\titlerunning{Latent multi-source correlation representation}
\author{Tongxue Zhou\inst{1} 
\and
St\'ephane Canu\inst{1}
\and 
Pierre Vera\inst{2}
\and 
Su Ruan\inst{1}}

\authorrunning{Z. Tongxue et al.}

\institute{
Normandie Univ, INSA Rouen, UNIROUEN, UNIHAVRE, LITIS, France \and
Department of Nuclear Medicine, Henri Becquerel Cancer Center, Rouen, 76038, France.\\
\email{tongxue.zhou@insa-rouen.fr}}
\maketitle              
\conf{\textbf{Accepted by 2020 23rd International Conference on Medical Image Computing \& Computer Assisted Intervention (MICCAI)\\
4-8 October, 2020, Lima, Peru}}

\begin{abstract}
Multimodal MR images can provide complementary information for accurate brain tumor segmentation. However, it's common to have missing imaging modalities in clinical practice. Since there exists a strong correlation between multi modalities, a novel correlation representation block is proposed to specially discover the latent multi-source correlation. Thanks to the obtained correlation representation, the segmentation becomes more robust in the case of missing modalities. The model parameter estimation module first maps the individual representation produced by each encoder to obtain independent parameters, then, under these parameters,the correlation expression module transforms all the individual representations to form a latent multi-source correlation representation. Finally, the correlation representations across modalities are fused via the attention mechanism into a shared representation to emphasize the most important features for segmentation. We evaluate our model on BraTS 2018 datasets, it outperforms the current state-of-the-art method and produces robust results when one or more modalities are missing.
\end{abstract}

\section{Introduction}
Brain tumor is one of the most aggressive and fatal cancers in the world, early diagnosis of brain tumors plays an important role in clinical assessment and treatment planning of brain tumors. Magnetic Resonance Imaging (MRI) is commonly used in radiology to diagnose brain tumors since it can provide complementary information due to its dependence on variable acquisition parameters, such as T1-weighted (T1), contrast-enhanced T1-weighted (T1c), T2-weighted (T2) and Fluid Attenuation Inversion Recovery (FLAIR) images. Different sequences can provide complementary information to analyze different subregions of gliomas. T2 and FLAIR are suitable to detect the tumor with peritumoral edema, while T1 and T1c are suitable to detect the tumor core without peritumoral edema~\cite{zhou2019review}. Therefore, applying multi-modal images can reduce the information uncertainty and improve clinical diagnosis and segmentation accuracy.

Segmentation of brain tumor by experts is expensive and time-consuming, recently, there have been many studies on automatic brain tumor segmentation\cite{cui2018automatic, zhao2018deep, havaei2017brain, wang2017automatic, kamnitsas2017efficient}, which always requires the complete set of the modalities. However, the imaging modalities are often incomplete or missing in clinical practice. Currently, there are a number of methods proposed to deal with the missing modalities in medical image segmentation, which can be broadly grouped into three categories: (1) training a model on all possible subset of the modalities, which is complicated and time-consuming. (2) synthesizing missing modalities and then use the complete imaging modalities to do the segmentation, while it requires an additional network for synthesis and the quality of the synthesis can directly affect the segmentation performance. (3) fusing the available modalities in a latent space to learn a shared feature representation, then project it to the segmentation space. This approach is more efficient than the first two methods, because it doesn't need to learn a number of possible subsets of the multi-modalities and will not be affected by the quality of the synthesized modality. Recently, there are a lot of segmentation methods based on exploiting latent feature representation for missing modalities. The current state-of-the-art network architecture is from Havaei, the proposed HeMIS~\cite{havaei2016hemis} learns the feature representation of each modality separately, and then the first and second moments are computed across individual modalities for estimating the final segmentation. However, computing mean and variance over individual representations can't learn the shared latent representation. Lau et al.~\cite{lau2019unified} introduced a unified representation network that maps a variable number of input modalities into a unified representation by using mean function for segmentation, while averaging the latent representations could averaging the latent representations could lose some important information. Chen et al.~\cite{chen2019robust} used feature disentanglement to decompose the input modalities into content code and appearance code, and then the content code are fused via a gating strategy into a shared representation for segmentation. While the approach is more complex and time-consuming, because it requires two encoders for each modalities, and their proposed fusion method only re-weight the content code from spatial-wise without considering the channel-wise. Shen et al.~\cite{shen2019brain} used adversarial loss to form a domain adaptation model to adapt feature maps from missing modalities to the one from full modalities, which can only cope with the one-missing modality situation. 

The challenge of segmentation on missing modalities is to learn a shared latent representation, which can take any subset of the image modalities and produce robust segmentation. To effectively learn the latent representation of individual representations, in this paper, we propose a novel brain tumor segmentation network to deal with the absence of imaging modalities. The main contributions of our method are three-fold: 1) A correlation representation block is introduced to discover the latent multi-source correlation representation. 2) A fusion strategy based on attention mechanism with obtained correlation representation is proposed to learn the weight maps along channel-wise and spatial-wise for different modalities. 3) The first multi-modal segmentation network which is capable of describing the latent multi-source correlation representation and allows to help segmentation for missing data is proposed. 

\section{Method}
\begin{figure}[htb]
\includegraphics[width=\textwidth]{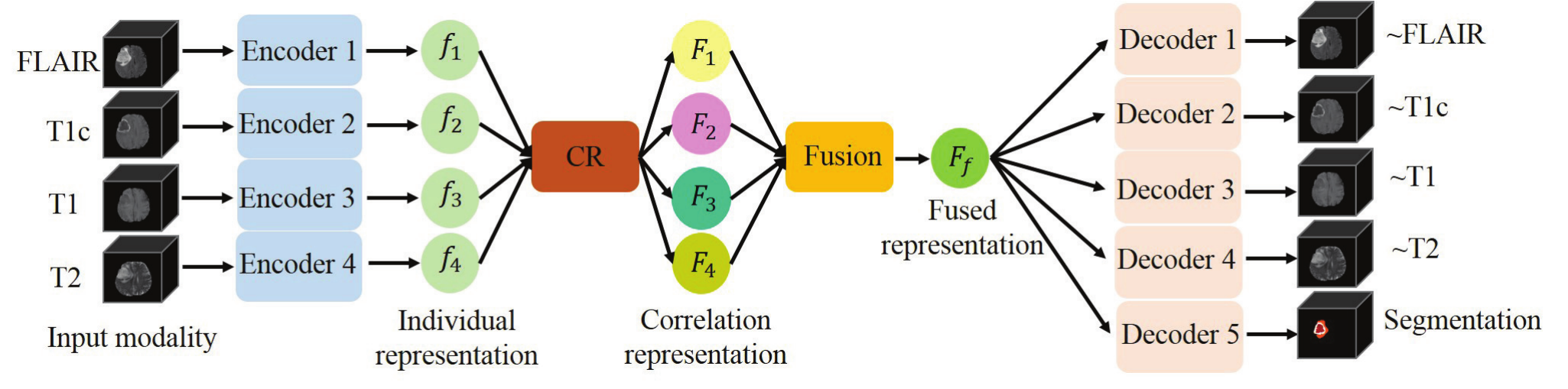}
\caption{A schematic overview of our network.}
\label{fig1}
\end{figure}

Our network is inspired by the U-Net architecture~\cite{ronneberger2015u}. To be robust to the absence of modalities, we adapt it to multi-encoder based framework. It first takes 3D available modalities as input in each encoder. The independent encoders can not only learn modality-specific feature representation, but also can avoid the false-adaptation between modalities. To take into account the strong correlation between multi modalities, we propose a block, named CR, to discover the correlation between modalities. Then the correlation representations across modalities are fused via attention mechanism, named Fusion, to emphasize the most discriminative representation for segmentation. Finally, the fused latent representation is decoded to form the final segmentation result. The network architecture scheme is depicted in Fig.~\ref{fig1}.

\subsection{Modeling the latent multi-source correlation} 
\begin{figure}[htb]
\centering
\includegraphics[width=0.8\textwidth]{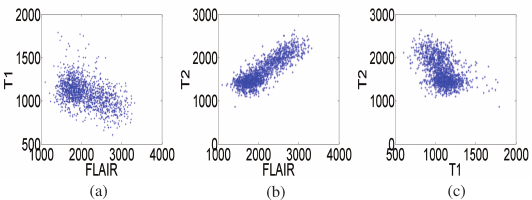}
\caption{Joint intensity distribution of MR images: (a) FLAIR-T1, (b) FLAIR-T2 and (c) T1-T2.}

\label{fig2}
\end{figure}

\begin{figure}[htb]
\centering
\includegraphics[width=0.9\textwidth]{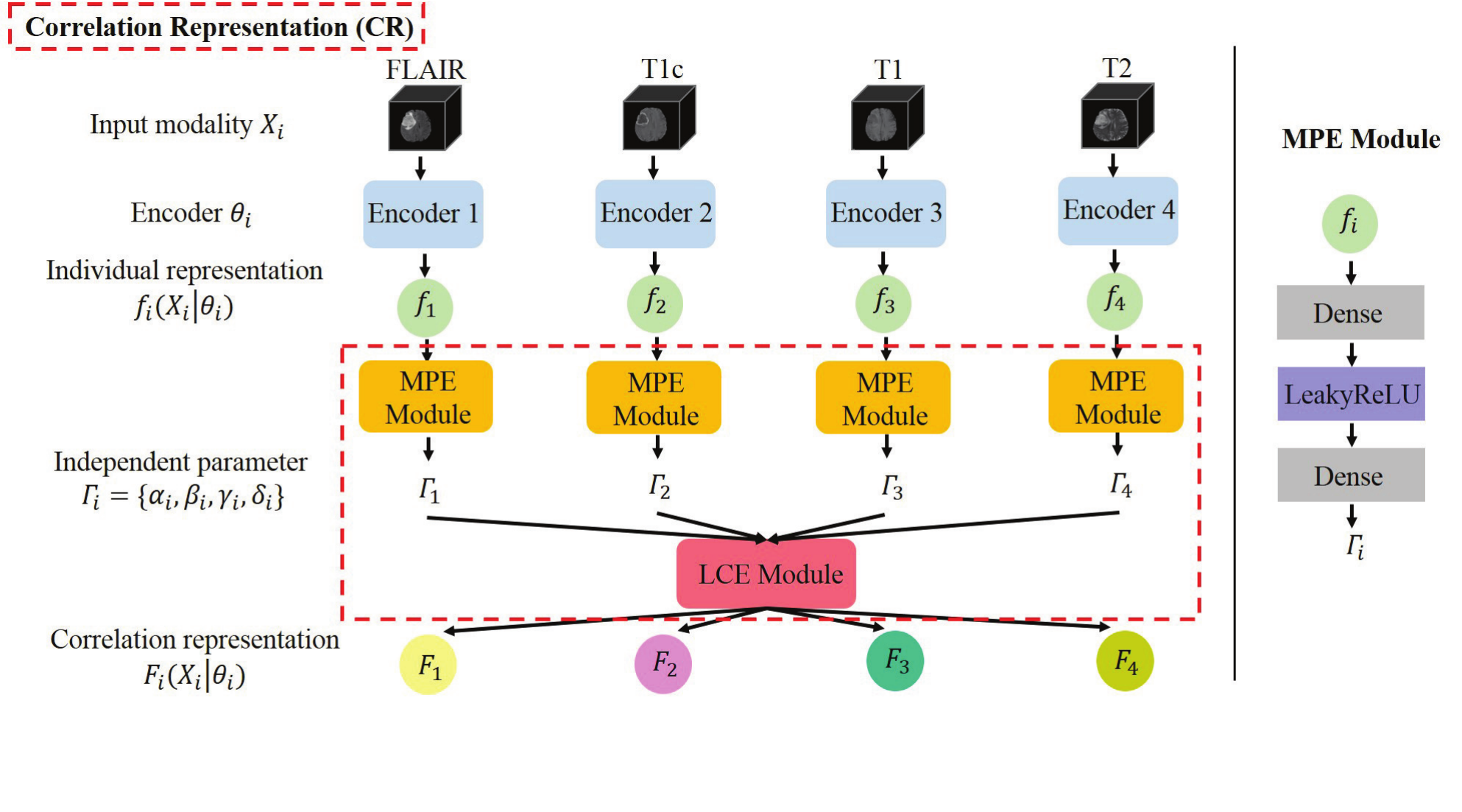}
\caption{Architecture of correlation representation block.}
\label{fig3}
\end{figure}

Inspired by a fact that, there is strong correlation between multi MR modalities, since the same scene (the same patient) is observed by different modalities~\cite{lapuyade2017segmenting}. From Fig.~\ref{fig2} presenting joint intensities of the MR images, we can observe a strong correlation in intensity distribution between each two modalities. To this end, it's reasonable to assume that a strong correlation also exists in latent representation between modalities. And we introduce a Correlation Representation (CR) block (see Fig.~\ref{fig3}) to discover the latent correlation. The CR block consists of two modules: Model Parameter Estimation Module (MPE Module) and Linear Correlation Expression Module (LCE Module). The input modality $\{X_i, ... , X_n\}$, where $n = 4$, is first input to the independent encoder $f_i$ (with learning parameters $\theta$) to learn the modality-specific representation $f_i(X_i|\theta_i)$. Then, MPE Module, a network with two fully connected network with LeakyReLU, maps the modality-specific representation $f_i(X_i|\theta_i)$ to a set of independent parameters $\Gamma_i =\{\alpha_i, \beta_i, \gamma_i, \delta_i\}$, which is unique for each moddality. Finally the correlation representation $F_i(X_i|\theta_i)$ can be obtained via LCE Module (Equation~\ref{eq1}). Since we have four modalities, we learn four correlations from the complete modalities. For the test, if one modality is missing, its feature representation can be approximately recovered from the learned correlation expression with the available modalities. We replace the missing modality by the most similar one to always have four inputs for the trained model. In this way, we do not lose the information of the missing modality for segmentation.

\begin{equation}
    F_i(X_i|\theta_i) = \alpha_i \odot f_j(X_{j}|\theta_{j})+\beta_i \odot f_k(X_{k}|\theta_{k})+\gamma_i \odot f_m(X_{m}|\theta_{m})+\delta_i,
    (i \neq j \neq k \neq m)
\label{eq1}
\end{equation}

\subsection{Fusion strategy}
\begin{figure}[htb]
\centering
\includegraphics[width=0.9\textwidth]{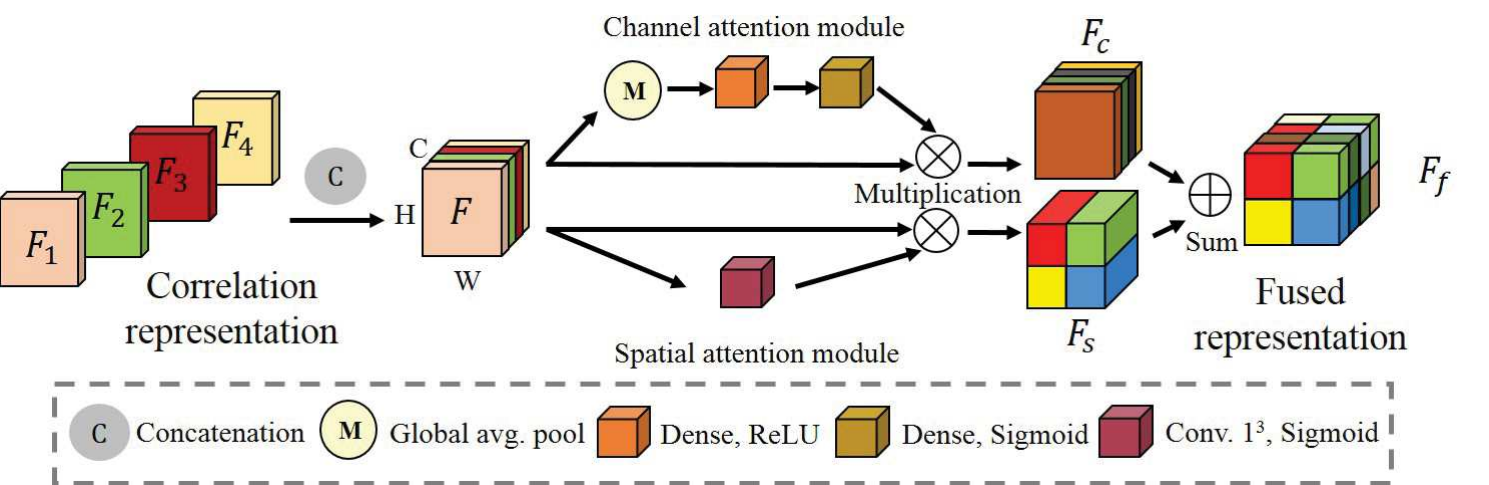}
\caption{The architecture of fusion block.}
\label{fig4}
\end{figure}

The proposed fusion strategy is based on two blocks: a CR block which searches for latent correlations between the feature representations of the four modalities and a fusion block, described in Fig.~\ref{fig4}, which seeks to weight the feature representations of the four modalities based on their contributions to the final segmentation. In our architecture, we use independently four encoders to obtain four feature representations corresponding to the four modalities. The CR block is then applied on the last layer of the encoders to learn the latent feature representation of each modality. Under our correlation hypothesis (Fig.~\ref{fig2}), each feature representation is linearly correlated with the other three modalities. Thus, the CR block learns four new representations with latent correlation. To learn the contributions of the four feature representations for the segmentation, we propose a fusion block based on attention mechanism, which allows to selectively emphasize feature representations. The fusion block consists of channel (modality) attention module and spatial attention module. The first module takes four feature representations as input to obtain the channel-wise weights. While the second module focuses on spatial location to obtain the spatial-wise weights. These two weights are combined with the input representation $F$ via multiplication to achieve the two attentional representations $F_c$ and $F_s$, which are finally added to obtain the fused representation $F_f$. The greater the weight of a modality, the greater the final contribution to its segmentation. In this way, we can discover the most relevant characteristics thanks to the fusion block, and recover the missing features thanks to the CR block, making the segmentation more robust in the case of the missing modalities. The proposed fusion block can be directly adapted to any multi modal fusion problem, and it encourages the network to learn more meaningful representation along spatial-wise and channel-wise, which is superior than simple mean or max fusion method.

\subsection{Network Architecture and Learning Process}
\begin{figure}[htb]
\centering
\includegraphics[width=0.9\textwidth]{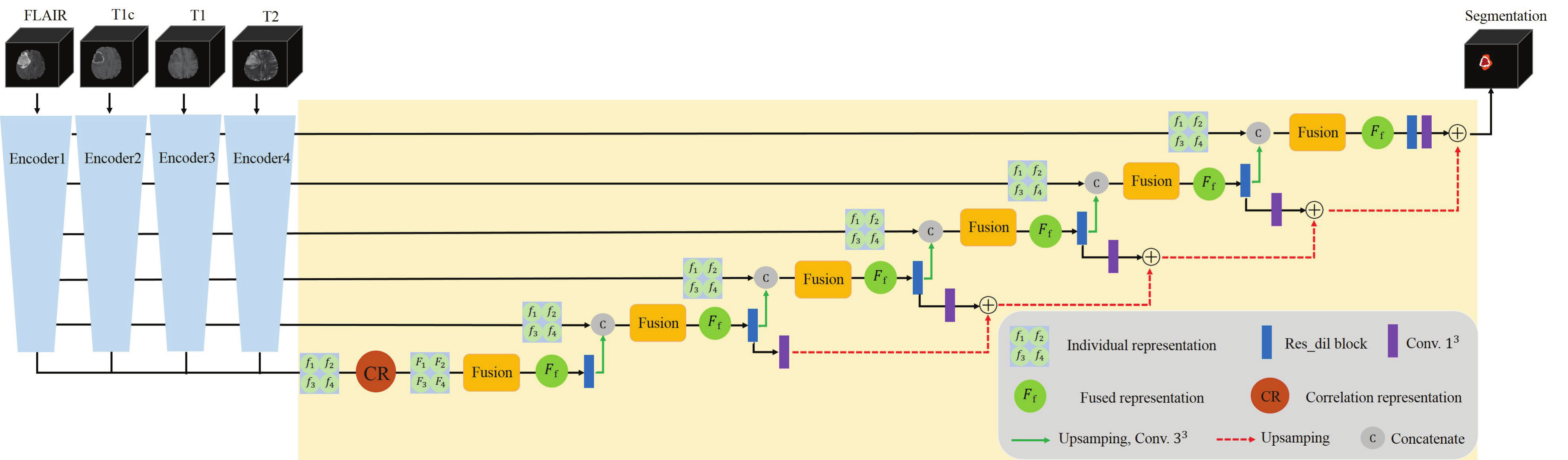}
\caption{Proposed segmentation network framework. Here only four encoders and the target segmentation decoder are shown.}
\label{fig5}
\end{figure}

The detailed network architecture framework is illustrated in Fig.~\ref{fig5}. It's likely to require different receptive fields when segmenting different regions in an image, standard U-Net can’t get enough semantic features due to the limited receptive field. Inspired by dilated convolution, we use residual block with dilated convolutions (rate= 2, 4) (res\_dil block) on both encoder part and decoder part to obtain features at multiple scale. The encoder includes a convolutional block, a res\_dil block followed by skip connection. All convolutions are $3\times3\times3$. Each decoder level begins with up-sampling layer followed by a convolution to reduce the number of features by a factor of 2. Then the upsampled features are combined with the features from the corresponding level of the encoder part using concatenation. After the concatenation, we use the res\_dil block to increase the receptive field. In addition, we employ deep supervision~\cite{isensee2017brain} for the segmentation decoder by integrating segmentation layers from different levels to form the final network output. The network is trained by the overall loss function: $\ L_{total}= L_{dice} + L_{1}$, where $L_1$ is the mean absolute loss.

\section{Data and implementation details}
\noindent\textbf{{Data and pre-processing.}}
The datasets used in the experiments come from BraTS 2018 dataset. The training set includes 285 patients, each patient has four image modalities including T1, T1c, T2 and FLAIR. Following the challenge, there are three segmentation classes: complete tumor, tumor core and enhancing tumor. The provided data have been pre-processed by organisers: co-registered to the same anatomical template, interpolated to the same resolution ($1 mm^3$) and skull-stripped. The ground truth have been manually labeled by experts. We did additional pre-processing with a standard procedure. The N4ITK~\cite{avants2009advanced} method is used to correct the distortion of MRI data, and intensity normalization is applied to normalize each modality of each patient. To exploit the spatial contextual information of the image, we use 3D image and resize it from $155 \times 240 \times240$ to $128 \times 128 \times128$. 

\noindent\textbf{{Implementation details.}}
Our network is implemented in Keras. The model is optimized using the Adam optimizer (initial learning rate = 5e-4) with a decreasing learning rate factor 0.5 with patience of 10 epochs in 50 epochs. We randomly split the dataset into 80\% training and 20\% testing. All the results are obtained by online evaluation platform.

\begin{table} 
\caption{Robust comparison of different methods (Dice \%) on BraTS 2018 dataset, $\bullet$ denotes the present modality and $\circ$ denotes the missing modality, $\uparrow$ denotes the improvement of CR block, bold results denotes the best score.}
\label{}
\begin{center}
\setlength{\tabcolsep}{1mm}{
\begin{tabular}{c|cccccccccccc}
\hline
{Modalities}
&\multicolumn{3}{c}{Complete}&\multicolumn{3}{c}{Core}&\multicolumn{3}{c}{Enhancing} \\
\hline
{\ F \ T1 \ T1c \ T2}
&\multicolumn{3}{c}{HeMIS \ Org \ Our}&\multicolumn{3}{c}{HeMIS \ Org \ Our}&\multicolumn{3}{c}{HeMIS \ Org \ Our}\\
\hline

$\circ$ \quad$\circ$ \quad$\circ$ \quad $\bullet$ & \ \textbf{38.6}& \ 31.4&  \ 32.3$\uparrow$& \ \textbf{19.5}& \ 14.9& \ 15.7$\uparrow$& \  0.0& \  6.2&  \ \textbf{7.2}$\uparrow$\\

$\circ$ \quad$\circ$ \quad$\bullet$ \quad $\circ$ &\ 2.6&\ 29.7&\ \textbf{33.5}$\uparrow$&6.5&49.3&\textbf{55.9}$\uparrow$&11.1&50.0&\textbf{53.5}$\uparrow$\\

$\circ$ \quad$\bullet$ \quad$\circ$ \quad $\circ$ &0.0&3.3&\textbf{5.3}$\uparrow$&0.0&4.3&\textbf{6.3}$\uparrow$&0.0&4.5&\textbf{5.3}$\uparrow$\\

$\bullet$ \quad$\circ$ \quad$\circ$ \quad $\circ$ &55.2&71.4&\textbf{73.7}$\uparrow$&16.2&46.2&\textbf{48.6}$\uparrow$&6.6&5.0&\textbf{25.8}$\uparrow$\\

$\circ$ \quad$\circ$ \quad$\bullet$ \quad $\bullet$ &48.2& 45.1&\textbf{48.3}$\uparrow$&45.8&48.1&\textbf{50.4}$\uparrow$&\textbf{55.8}&52.0&52.4$\uparrow$\\

$\circ$ \quad$\bullet$ \quad$\bullet$ \quad $\circ$ &15.4&11.4&\textbf{29.2}$\uparrow$&30.4&22.6&\textbf{54.8}$\uparrow$&42.6&24.8&\textbf{53.8}$\uparrow$\\

$\bullet$ \quad$\bullet$ \quad$\circ$ \quad $\circ$ &71.1&75.9&\textbf{80.4}$\uparrow$&11.9&47.4&\textbf{51.5}$\uparrow$&1.2&7.7&\textbf{10.2}$\uparrow$\\

$\circ$ \quad$\bullet$ \quad$\circ$ \quad $\bullet$ &\textbf{47.3}&31.6&35.5$\uparrow$&\textbf{17.2}&12.9&14.3$\uparrow$&0.6&2.5&\textbf{6.1}$\uparrow$\\

$\bullet$ \quad$\circ$ \quad$\circ$ \quad $\bullet$ &74.8&80.4&\textbf{81.3}$\uparrow$&17.7&20.7&\textbf{25.0}$\uparrow$&0.8&9.3&\textbf{10.0}$\uparrow$\\

$\bullet$ \quad$\circ$ \quad$\bullet$ \quad $\circ$ &68.4&80.3&\textbf{81.5}$\uparrow$&41.4&65.7&\textbf{73.4}$\uparrow$&53.8&62.7&\textbf{67.5}$\uparrow$\\

$\bullet$ \quad$\bullet$ \quad$\bullet$ \quad $\circ$ &70.2&81.1&\textbf{82.7}$\uparrow$&48.8&71.7&\textbf{75.8}$\uparrow$&60.9&65.7&\textbf{68.4}$\uparrow$\\

$\bullet$ \quad$\bullet$ \quad$\circ$ \quad $\bullet$ &75.2&83.5&\textbf{85.4}$\uparrow$&18.7&41.3&\textbf{44.4}$\uparrow$&1.0&11.1&\textbf{12.9}$\uparrow$\\

$\bullet$ \quad$\circ$ \quad$\bullet$ \quad $\bullet$ &75.6&87.5&\textbf{87.7}$\uparrow$&54.9&74.2&\textbf{77.4}$\uparrow$&60.5&65.4&\textbf{67.2}$\uparrow$\\

$\circ$ \quad$\bullet$ \quad$\bullet$ \quad $\bullet$ &44.2&46.9&\textbf{50.1}$\uparrow$&46.6&51.2&\textbf{52.1}$\uparrow$&\textbf{55.1}&54.3&54.8$\uparrow$\\

$\bullet$ \quad$\bullet$ \quad$\bullet$ \quad $\bullet$ &73.8&87.9&\textbf{88.1}$\uparrow$&55.3&76.2&\textbf{78.8}$\uparrow$&61.1&68.1&\textbf{69.1}$\uparrow$\\

\hline
Wins / 15 &2&0&13&2&0&13&2&0&13\\
\hline
\end{tabular}
\label{tab1}}
\end{center}
\end{table}

\section{Experiments Results}
\noindent\textbf{{Quantitative Analysis.}}
The main advantage of our method is using the correlation representation, which can discover the latent correlation representation between modalities to make the model robust at the absence of modalities. To prove the effectiveness of our model, we use Dice score as the metric, and compare two other approaches. (1) HeMIS~\cite{havaei2016hemis}, the current state-of-the-art method for segmentation with missing modalities. Since the original work didn't publish the available code, the reported results on HeMIS are taken from the work in \cite{dorent2019hetero}. (2) Org, a specific case of our model without correlation representation block. From Table~\ref{tab1}, for all the tumor regions, our method achieves the best results in most of all cases. Compared to HeMIS, the Dice score of our method just gradually drops when modalities are missing, while the performance drop is more severe in HeMIS. Compared to Org, the correlation representation block makes the model more robust in the case of missing modalities, which demonstrates the effectiveness of the proposed component and also proves our assumption. We can also find that, missing FLAIR modality leads to a sharp decreasing on dice score for all the regions, since FLAIR is the principle modality for showing whole tumor. Missing T1 and T2 modalities would contribute to a slight decreasing on dice score for all the regions. While missing T1c modality would contribute to a sever decreasing on dice score for both tumor core and enhancing tumor, since T1c is the principle modality for showing tumor core and enhancing tumor regions.

\begin{figure}[htb]
\centering
\includegraphics[width=\textwidth]{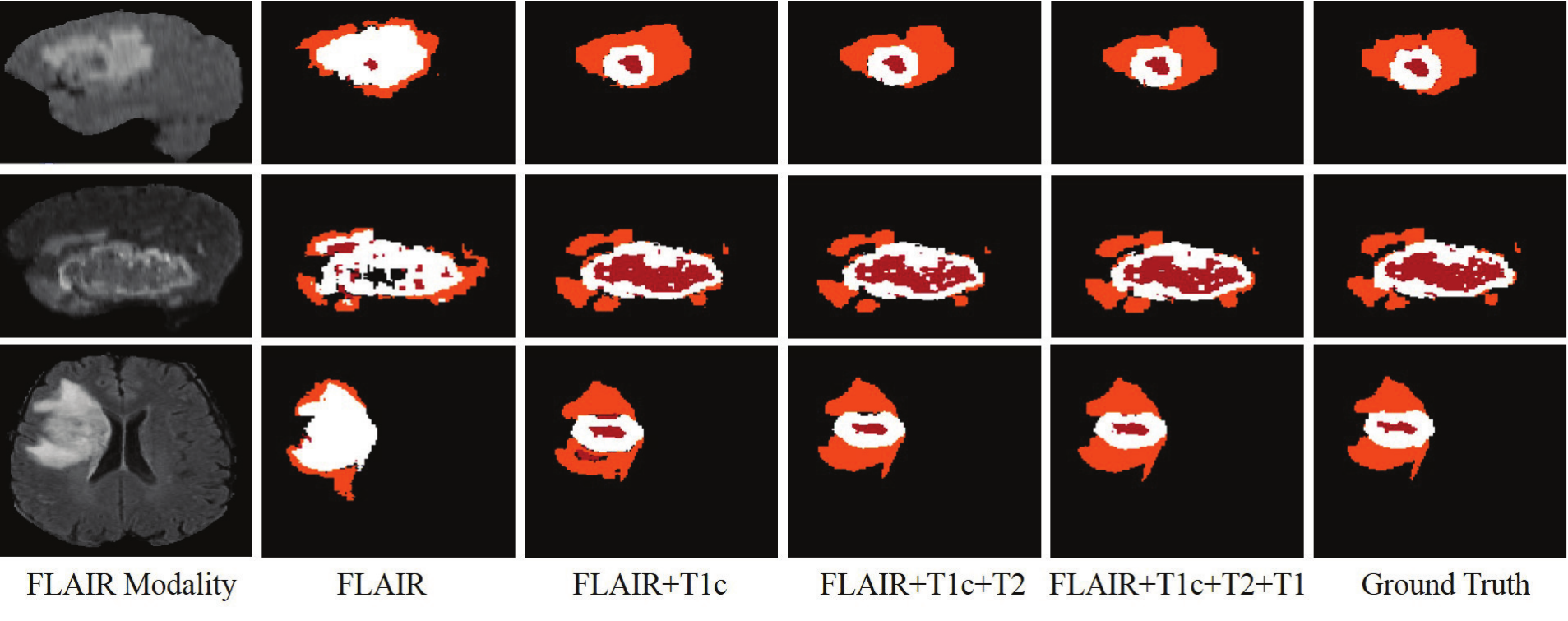}
\caption{Examples of the segmentation results compared to ground truth. Red: necrotic and non-enhancing tumor core; Orange: edema; White: enhancing tumor.}
\label{fig6}
\end{figure}

\noindent\textbf{{Qualitative Analysis.}}
In order to evaluate the robustness of our model, we randomly select three examples on BraTS 2018 dataset and visualize the segmentation results in Fig.~\ref{fig6}. We can observe that with the increasing number of missing modalities, the segmentation results produced by our robust model just slightly degrade, rather than a sudden sharp degrading. Even with FLAIR and T1c modalities, we can achieve a decent segmentation result.

\section{Conclusion}
We propose a novel multimodal brain tumor segmentation network based on the latent multi-source correlation representation and fusion using attention mechanism for making the model robust to missing data. We demonstrate our method can yield competitive results on BraTS 2018 dataset under both full and missing modalities. The comparison results also show that the important roles of FLAIR and T1c on segmenting the complete tumor and tumor core, respectively. The proposed method can be generalized to other segmentation tasks with other modalities (e.g. MR and CT images). In the future we will test our method on other segmentation datasets and compare with other latent representation learning methods\cite{chartsias2017multimodal, chartsias2019multimodal}. In addition, we will investigate more complex model to describe the multi-source correlation representation and adapt it to missing data issue.
\bibliographystyle{splncs04}
\bibliography{paper2409}

\end{document}